\documentclass[twocolumn]{aastex63}

\newcommand{\oii}[1]{[{\sc O\,ii}]}
\newcommand{\oiii}[1]{[{\sc O\,iii}]}
\newcommand{\nii}[1]{[{\sc N\,ii}]}
\newcommand{\sii}[1]{[{\sc S\,ii}]}

\newcommand{\hii}[1]{{\sc H\,ii}}

\newcommand\carter[1]{{\color{teal} #1}}

\newcommand\kms{$\mathrm{km\,s}^{-1}\,$}
\received{}
\revised{}
\accepted{\today}
\submitjournal{ApJ}

\usepackage{subfig}

\shorttitle{Machine Learning Approach to Spectroscopic Observations II}
\shortauthors{Rhea et al.}
\usepackage{amsmath}

\DeclareMathOperator*{\argmin}{arg\,min}
\graphicspath{{./}{figures/}}
\usepackage{comment}
\begin{document}

\title{A Machine Learning Approach to Integral Field Unit Spectroscopy Observations: II. HII Region Line Ratios}

\correspondingauthor{Carter Rhea}
\email{carterrhea@astro.umontreal.ca}

\author[0000-0003-2001-1076]{Carter Rhea}
\affiliation{D\'epartement de Physique, Universit\'e de Montr\'eal, Succ. Centre-Ville, Montr\'eal, Qu\'ebec, H3C 3J7, Canada}

\author[0000-0002-5136-6673]{Laurie Rousseau-Nepton}
\affiliation{Canada-France-Hawaii Telescope, Kamuela, HI, United
States}

\author[0000-0002-1755-4582]{Simon Prunet}
\affiliation{Canada-France-Hawaii Telescope, Kamuela, HI, United
States}
\affiliation{Laboratoire Lagrange, Universit\'e C\^ote d’Azur, Observatoire de la C\^ote d’Azur, CNRS, Parc Valrose, 06104 Nice Cedex 2, France}

\author[0000-0002-2457-3431]{Myriam Prasow-Émond}
\affiliation{D\'epartement de Physique, Universit\'e de Montr\'eal, Succ. Centre-Ville, Montr\'eal, Qu\'ebec, H3C 3J7, Canada}

\author[0000-0001-7271-7340]{Julie Hlavacek-Larrondo}
\affiliation{D\'epartement de Physique, Universit\'e de Montr\'eal, Succ. Centre-Ville, Montr\'eal, Qu\'ebec, H3C 3J7, Canada}

\author[0000-0003-0842-8688]{Natalia Vale Asari}
\affiliation{Departamento de F\'{\i}sica--CFM, Universidade Federal de Santa Catarina, C.P.\ 476, 88040-900, Florian\'opolis, SC, Brazil}
\affiliation{School of Physics and Astronomy, University of St Andrews, North Haugh, St Andrews KY16 9SS, UK}
\affiliation{Royal Society--Newton Advanced Fellowship }

\author[0000-0002-3247-5321]{Kathryn Grasha}
\affiliation{Research School of Astronomy and Astrophysics, Australian National University, Weston Creek, ACT 2611, Australia}

\author[0000-0003-3544-3939]{Laurence Perreault-Levasseur}
\affiliation{D\'epartement de Physique, Universit\'e de Montr\'eal, Succ. Centre-Ville, Montr\'eal, Qu\'ebec, H3C 3J7, Canada}
\affiliation{Mila - Quebec Artificial Intelligence Institute, Montreal, Qu\'ebec, Canada}
\affiliation{Center for Computational Astrophysics, Flatiron Institute, New York, USA}

\begin{abstract}
In the first paper of this series (\citealt{rhea_machine_2020}), we demonstrated that neural networks can robustly and efficiently estimate kinematic parameters for optical emission-line spectra taken by SITELLE at the Canada-France-Hawaii Telescope. This paper expands upon this notion by developing an artificial neural network to estimate the line ratios of strong emission-lines present in the SN1, SN2, and SN3 filters of SITELLE. We construct a set of 50,000 synthetic spectra using line ratios taken from the Mexican Million Model database replicating \hii{} regions. Residual analysis of the network on the test set reveals the network's ability to apply tight constraints to the line ratios. We verified the network's efficacy by constructing an activation map, checking the \nii{} doublet fixed ratio, and applying a standard k-fold cross-correlation.
Additionally, we apply the network to SITELLE observations of M33; the residuals between the algorithm's estimates and values calculated using standard fitting methods show general agreement. Moreover, the neural network reduces the computational costs by two orders of magnitude. Although standard fitting routines do consistently well depending on the signal-to-noise ratio of the spectral features, the neural network can also excels at predictions in the low signal-to-noise regime within the controlled environment of the training set as well as on observed data when the source spectral  properties are well constrained by models. 
These results reinforce the power of machine learning in spectral analysis. 
\end{abstract}

\section{Introduction}\label{sec:intro}
Emission-line nebulae are a critical part of our understanding of galactic evolution and radiative processes; thus, they are a primary targets of observation in extragalactic astronomy (\citealt{kennicutt_structural_1984}; \citealt{veilleux_spectral_1987}; \citealt{kewley_understanding_2019}). 
\hii{} regions form from clumps of gas in the interstellar medium (ISM) when young O/B stars irradiate the surrounding environment (e.g. \citealt{franco_evolution_2000}; \citealt{osterbrock_astrophysics_1989}; \citealt{shields_extragalactic_1990}). 
The region becomes either partially or fully ionized depending on the budget and hardness of ionizing photons, the morphology and the total mass of the mother cloud. \hii{} regions are primarily composed of Hydrogen and Helium; however, they contain non-negligible amounts of metals (e.g. \citealt{shields_composition_1976}; \citealt{garnett_composition_1987}; \citealt{kennicutt_systematic_1993}; \citealt{oey_abundances_1993}). 
Through recombination and collisonal processes between the ionized atoms and the free electrons, the nebula emits characteristic strong emission-lines which indicate their underlying chemical structure (e.g. \citealt{baldwin_classification_1981}; \citealt{crawford_rosat_1999}; \citealt{kewley_optical_2001}; \citealt{kewley_host_2006};  \citealt{kewley_understanding_2019}).
In the optical, the primary emission lines include, but are not limited to, the Balmer series 
(i.e. H\,$\alpha\lambda$6563, and H\,$\beta\lambda$4861), ionized oxygen (i.e. \oii{}$\lambda\lambda$3726,\,3729, and \oiii{}$\lambda\lambda$4959,\,5007), ionized nitrogen (i.e. \nii{}$\lambda\lambda$6548,\,6583), and ionized sulfur (i.e. \sii{}$\lambda\lambda$6716,\,6731).
By studying the relative ratios of these lines, the primary emission and ionization mechanism can be determined.
Sources that power ionized emission differ and the regions themselves are categorized into several distinct classes: classical \hii{} regions (e.g. \citealt{viallefond_star_1985}; \citealt{melnick_giant_1987}), shock induced regions such as supernova remnants (e.g. \citealt{fesen_optical_1985}; \citealt{danziger_optical_1976}), and planetary nebulae (e.g. \citealt{miller_planetary_1974}; \citealt{oserbrock_planetary_1964}). Several diagnostics exist to categorize emission nebulae based off their line ratios (e.g. \citealt{baldwin_classification_1981}; \citealt{kewley_understanding_2019};  \citealt{constantin_clustering_2006}; \citealt{dagostino_new_2019}). Nevertheless, these different diagnostics have limitations due to the multi-parameters physical nature of these objects which is not exempt of degeneracy.  
With that in mind, we want to test the hypothesis that fitting multiple strong lines directly using model predictions (e.g. CLOUDY, MAPPINGS; \citealt{ferland_2017_2017} \citealt{allen_mappings_2008}) that covers typical physical conditions in the gas could help minimize the errors associated on the line parameters (intensity, broadening and velocity), provide a pathway to classification of the nebulae directly from the fit, and potentially significantly increase the computational efficiency of the procedure as well as providing a new angle to estimate uncertainties.

The recent advent of integral field spectroscopy is expanding our knowledge of emission-line nebulae with their increased spatial and spectral resolution; these instruments are pushing the ability of existing analysis tools due to their resolution.  (e.g. \citealt{sanchez_integral_2012}; \citealt{leroy_portrait_2016}; \citealt{bundy_overview_2014}; \citealt{martins_near-ir_2010}).
SITELLE, an Imaging Fourier Transform Spectrograph (IFTS) located at the Canada-France-Hawai'i Telescope, is one such instrument. SITELLE has an unrivaled field of view of 11$'\times$11$'$ and produces data cubes containing over 4 million pixels. Each pixel contains a spectrum.
(e.g. \citealt{baril_commissioning_2016}; \citealt{drissen_sitelle_2019}). SITELLE has a spectral resolution between 1 and 20,000. It's instrumental line shape is described by a cardinal sinc function. It can be convolved with a Gaussian (\citealt{martin_optimal_2016}) to account for natural broadening of the observed lines. Spectral fits must be done cautiously to accurately model the line shape and capture the effects of sidelobes of the sinc, which can greatly influence the estimated flux of an emission-line (e.g. \citealt{martin_orbs_2012}). In many field of astrophysics, it is necessary to model resolved and unresolved (blended) lines accurately. Emission line parameters are the root information used to infer physical properties of locally-resolved nebulae as well as classifying them, but they are also crucial for proper characterisation of galaxies at different redshift. For the later, the line parameters are essential to fully understand how internal and external processes affect galaxies as one cannot disentangle active galactic nucleus, stellar winds, supernovae, etc, multiple feedback processes that affect the whole galaxy in different ways and at different scales.  

In this paper we apply an artificial neural network to SITELLE spectra in order to obtain the strong emission-line ratios and validates its results by comparing with the line ratio derived from a standard line fitting techniques. We demonstrate the capability of the machine learning algorithm in low signal-to-noise regimes. In $\S$2, we describe the synthetic data used to train and test the algorithm. It is important to note that although the method is tested here on well resolved objects, it could as well be used on galaxy at higher redshift and on integrated spectra. Additionally, we dissect the neural network and its hyperparameters. In $\S$3, we describe how the network was trained and how it compares with the standard fitting procedure. In $\S$4, we explore the impact of the signal-to-noise on the algorithm in addition to applying it on a real observation of M33, a well resolved local galaxy. We conclude in $\S$5.

\section{Methodology} \label{sec:meth}

In the first article of the series, \citet{rhea_machine_2020}, we explored the application of a convolutional neural network to calculate the kinematic parameters from SITELLE spectra. In this paper we expand the use of machine learning to calculate another critical set of physical parameters, the line ratios of strong emission-lines. As in the previous paper, the initial step in any machine learning application is to assemble the appropriate training set.

\subsection{Synthetic Data}\label{sec:syn}
In order to facilitate the training of the network used to estimate strong-line ratios, we rely on carefully constructed synthetic spectral data. Synthetic data for emission-line sources must account for both the instrumental line shape (ILS) of an instrument for a given spectral resolution and the relative intensities and broadening of the lines available through observations. 
The network was developed for use in the SIGNALS program, which uses SITELLE to observe in three bands: SN1 (365--385 nm), SN2 (480--520 nm), and SN3 (651--685 nm). The synthetic data, and observed data, contain approximately 841 channels in SN3, 219 channels in SN2, and 171 channels in SN1.
Each strong emission-line's ILS (H$\alpha$(6563)\AA, \oii{}$\lambda$3726, \oii{}$\lambda$3729, H$\beta$(4861)\AA, \oiii{}$\lambda$4959, \oiii{}$\lambda$5007, [{\sc N\,ii}]$\lambda$6548, [{\sc N\,ii}]$\lambda$6583,
 (H$\alpha$(6563)\AA, [{\sc S\,ii}]$\lambda$6716, and [{\sc S\,ii}]$\lambda$6731,  was separately modeled using the routine \texttt{orb.fit.create\_cm1\_lines\_model} (\citealt{martin_orbs_2012}); \texttt{ORB} is the software kernel written for SITELLE, the imaging Fourier Transform Spectrometer on the Canada-France-Hawaii Telescope (\citealt{martin_optimal_2016}; \citealt{martin_calibrations_2017}; \citealt{baril_commissioning_2016}). \texttt{ORBS} and \texttt{ORCS} build upon the \texttt{ORB} kernal and are used for data reduction and data analysis, respectively.

 Following the SIGNALS large program instrumental configuration for the observations, we set the maximal spectral resolution in SN1 and SN2 to R$\sim$1000 and in SN3 to R$\sim$5000 (\citealt{rousseau-nepton_signals_2019}) to create the training set. Since the resolution varies as a function of location in a SITELLE spectral cube, we allow for variations within our spectral resolution to R$\sim$200 less than the maximal value (e.g. \citealt{martin_calibrations_2017}). Previous work reveled that this is a key parameter for an accurate training of the algorithm (\citealt{rhea_machine_2020}).
To fully sample the velocity and broadening space expected in the SIGNALS catalog (and in keeping with our previous study), we randomly select the velocity parameter from a uniform distribution between -200 and 200 \kms 
and the broadening parameter between 10 and 50 \kms. 
These values represent expected values for typical HII regions (e.g. \citealt{epinat_ghasp_2008}); additionally, at R$\sim$5000 SITELLE is unable to resolve the broadening parameter below 10 \kms (\citealt{rousseau-nepton_signals_2019}). 
The resolution, broadening, and velocity were randomly selected with replacement for each synthetic spectra so that we sample the entire parameter space (e.g., \citealt{james_introduction_2013}).
Furthermore, the signal-to-noise ratio varied randomly between 5 and 30 with respect to the Halpha emission -- meaning that Halpha emission is at least 5 times over the noise. However, this does not ensure that all lines are above the noise. The noise was assigned to each spectral channel individually and was randomly sampled from a normal distribution centered around the chosen signal-to-noise ratio with a sigma of 1.

The last element required to create synthetic spectra is the relative amplitude of the strong emission-lines. For each type of nebulae, we used different databases. Following the methodology described in detail in \cite{rhea_machine_2020}, relative line amplitude of the HII regions were sampled from the Mexican Million Models database (3MdB; \citealp{morisset_virtual_2015}) BOND simulations \citep{asari_bond_2016}. 
Following standard procedure, we use 70\% of the synthetic data for the training set, 20\% for the validation set, and 10\% for the test set (e.g., \citealt{breiman_random_2001}). 

In contrast with our previous study, \citealt{rhea_machine_2020}, we require all strong lines sampled in the SIGNALS observations to be present. We also add the restriction that all strong emission-lines must have an amplitude equal to or greater than 12\% of the H$\alpha$ amplitude. This threshold corresponds to a signal-to-noise ratio of 3 for the faintest targets in the SIGNALS sample whose H$\alpha$ surface brightness is approximately $8\times10^{-17}$ erg s$^{-1}$ cm$^{-2}$ arcsec$^{-2}$ (\citealt{rousseau-nepton_signals_2019}).
While the impact of this threshold is discussed more in-depth in $\S$ \ref{sec:snr}, note that consistently, the training, validation, and test sets for the primary CNN used in this paper all adhere to this signal-to-noise constraint. Moreover, we select a Balmer decrement by randomly sampling a value between 2 and 6 from a uniform distribution with a sampling increment of 0.01. We then apply reddening at each wavelength sampled in our spectra following the procedure outlined in $\S$ \ref{sec:red}. 

We create 50,000 mock spectra in the form of FITS files which contain the emission-line information  (e.g. velocity, broadening, resolution, emission-line fluxes) in addition to the spectrum itself.

\begin{table*}[hbt]
    \centering
    \begin{tabular}{|c|c|c|c|c|c|c|c|}
         \hline
         [{\sc S\,ii}]$\lambda6731$/[{\sc S\,ii}]$\lambda6716$ & [{\sc S\,ii}]/H$\alpha$  & [{\sc N\,ii}]/H$\alpha$ & [{\sc N\,ii}]/[{\sc S\,ii}] & [{\sc O\,iii}]/H$\beta$ & [{\sc O\,ii}]/H$\beta$ & [{\sc O\,ii}]/[{\sc O\,iii}] & H$\alpha$/H$\beta$* \\
         \hline \hline 
          0.73-0.79 & 0.27-1.18 & 0.27-2.03 & 0.14-6.97 & 0.33-2.47 & 0.24-6.97 & 0.60-30.89 & 1.93-5.99\\
         \hline
    \end{tabular}
    \caption{Ranges of line ratio parameters from synthetic spectra. The distribution of the parameters are all skewed heavily to the lower values (with the exception of the \sii{} doublet ratio which is evenly distributed over a small range) which represents the likelihood of finding those parameters in the simulations from which these values are obtained. After applying an \texttt{arcsinh} transformation, the variables are normally distributed.
    * The wide range is due to artificially injected dust redenning following the prescription described in $\S$\ref{sec:red}.}
    \label{tab:Dataset}
\end{table*}

\subsection{Artificial Neural Networks} \label{sec:ann}
In this paper, we study the application of an artificial neural network to the problem of strong emission-line ratio estimation. Using the synthetic spectra described in $\S$\ref{sec:syn}, we train a network to approximate the function which maps SITELLE spectra to their corresponding line ratios. 
\subsubsection{The Algorithm}
Feed Forward artificial neural networks (ANN) contain three principal layers: the input layer, the hidden layer(s), and the output layer (e.g. \citealt{hansen_neural_1990}). The input layer consists of the preproccessed data that will be used to train the network and eventually be fed unseen inputs to make predictions. In this case, the input layer is the combined SN1, SN2, and SN3 SITELLE observations described previously. The output layer consists of line ratio estimates. The hidden layers contain an ensemble of nodes which are parametrized by a linear function that takes the input, \textbf{x}, multiplies it by a weight matrix, \textbf{w}, and adds a bias, \textbf{b}. The node is then activated by a predetermined function. A common activation function, the Rectified Linear Unit (ReLU), is employed in every layer of the network used in this paper except for the final layer (e.g. \citealt{chen_non-linear_1990}). The ReLU function, $g(w\times x+b)=\max(0,w\times x+b)$, takes the value of the node unless the value is negative; in that case, it takes the value 0. The result is a non-negative value for each node in a layer where the vector-valued function of each layer, $l$, is denoted as \textbf{h}$^l$(\textbf{x},\textbf{b},\textbf{w}). In traditional neural networks, such as the network applied in this work, the layers are \textit{fully-connected}; thus, each node in a layer is connected to all nodes in the previous and subsequent layer. 

After calculating \textbf{h}$^l$(\textbf{x},\textbf{b},\textbf{w}) sequentially for each layer, we have a vector-valued output \textbf{f}(\textbf{x},\textbf{b},\textbf{w}). We can calculate the \textit{loss} between the final predictions, \textbf{f}, and the correct outputs, \textbf{y}. 
We adopt the Huber loss function which is defined as:
\begin{equation}
    k_\delta(f,y:w,b)) =
    \begin{cases} 
      \frac{1}{2}(y-f(w,b))^2 & |y-f(w,b)|\leq \delta \\
      \delta|y-f(w,b)|-\frac{1}{2}\delta^2 & \text{otherwise}
   \end{cases}
\end{equation}
where $\delta$ is a tune-able parameter and initialized as 1. The Huber loss function is often employed since it reduces the effects of outliers on the final cost calculation (e.g. \citealt{huber_robust_1964}).

We then minimize the loss function through back-propagation in which we alter the weights and biases (\citealt{hecht-nielsen_theory_1989}):
\begin{equation}
    \argmin_{w,b}k(f,y; w,b)
\end{equation}
We apply the Adam implementation of the stochastic gradient descent algorithm in order to minimize the loss function (\citealt{lechevallier_proceedings_2010}; \citealt{kingma_adam_2017}). In this manner, we are able to train the network by updating the weights and biases until the loss function is minimized on the training set.

\subsection{The Network and Hyper-Parameters}
\begin{figure}
    \centering
    \includegraphics[width=0.48\textwidth]{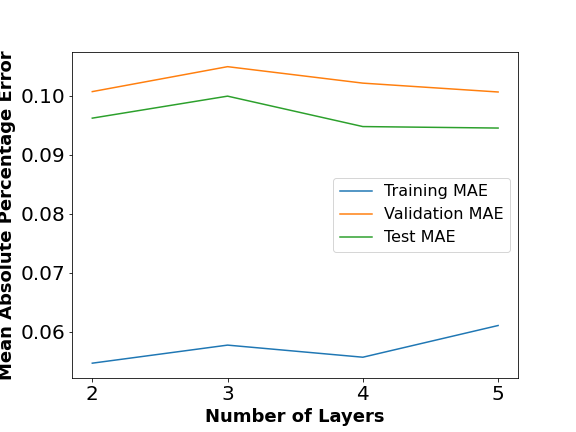}
    \caption{Final mean absolute percentage error after five epochs of training as a function of the number of layers in the neural network. Each layer contains 1000 nodes and is activated by a \texttt{relu} function. The graphics show the curve for the mock training, validation, and test sets. We emphasize that the synthetic data used in the structural parameter tuning were not used again.}
    \label{fig:NetworkLayers}
\end{figure}

\begin{figure*}
    \centering
    \includegraphics[width=0.9\textwidth]{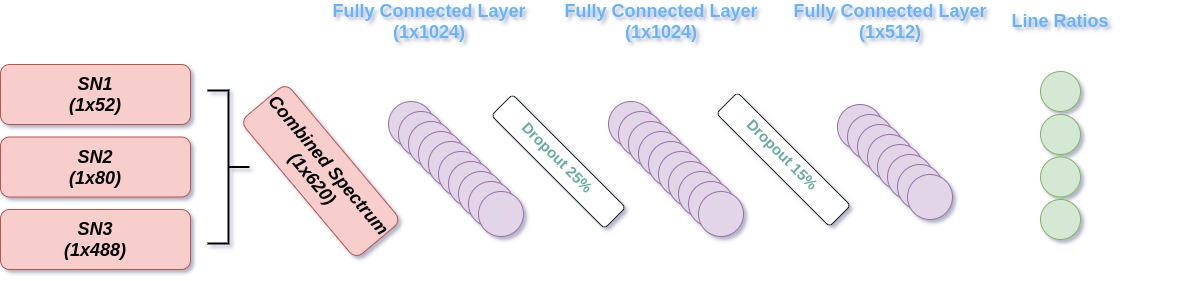}
    \caption{Graphical depiction of the artificial neural network employed in this paper. The spectrum vectors for SN1, SN2, and SN3 are combined to form a single input vector which is then passed through a series of three fully connected layers. A linear activation function takes the final layer and compresses it into a final prediction for each line ratio.}
    \label{fig:Network}
\end{figure*}

In order to determine the structural parameters of the network, we first constructed four networks; The first network had two hidden layers, the second network three hidden layers, and so on. Since the problem is nonlinear, we did not test a single hidden layer network. We note that the training, validation, and test sets used to determine the structural parameters contain only 10000 synthetic spectra. These spectra were not re-used in the final training, validation, and testing ensemble. Initially, each hidden layer contained 1000 nodes. This number is sufficient to test the optimal number of layers (e.g. \citealt{sheela_review_2013}). 
Figure \ref{fig:NetworkLayers} demonstrates that the MAPE plateaus for the training set at four layers. However, since the mean squared error (MSE) increase in both the validation and test sets between three and four layers, we chose three layers as the optimal value. Although the figure is not included, the same patterns hold for the mean absolute error. 
Next, we applied a grid search on the number of hidden neurons allowing each to vary from 64, 128, 256, 512, and 1024 nodes. The MSE was minimized for all sets when the first layer has 1024 nodes, the second layer has 1024 nodes, and the third layer has 512 nodes.
Figure \ref{fig:Network} graphically depicts the entire network. The structure of the algorithm is as follows:
\begin{enumerate}
    \item Concatenate spectral vectors from SN1, SN2, and SN3 as input
    \item Fully Connected Layer with ReLU Activation (1x1024)
    \item Dropout (25\%)
    \item Fully Connected Layer with ReLU Activation (1x1024)
    \item Dropout (15\%)
    \item Fully Connected Layer with ReLu Activation (1x512)
    \item Fully Connected Output Layer with Linear Activation (1x8)
\end{enumerate}

We determined the optimal hyperparameter values by using the grid search and random search cross-validation techniques implemented in \texttt{sklearn}.
The batch size's optimal value was determined to be four.  Additionally, we adopt a normal random distribution as the initialization of the hidden layer neurons (\citealt{thimm_neural_1995}; \citealt{de_castro_feedforward_1998}). The data are normalized using to the maximum value in the the SN3 filter which corresponds to the maximum value in the combined filter. We note that this is not necessarily the amplitude of the H$\alpha$ line. We use the \texttt{tensorflow} implementation of the \texttt{ADAM} optimization algorithm.

Initial testing of the algorithm revealed a systematic bias of approximately 10\% in the residuals (see $\S$\ref{sec:res} for more details). In order to account for the bias, we applied two techniques: a transformation of the target variables and a regularization parameter. Initially, the distributions of the target variables (i.e. the line ratios the network is predicting) were skewed positive; we thus applied several transforms on the target variables (e.g. \texttt{log10}, \texttt{ln}, \texttt{arcsinh}, etc.). Experimentation showed that applying an \texttt{arcsinh} transformation resulted in normalized target variable distributions. Furthermore, this reduced the systematic bias considerably (e.g. \citealt{zheng_feature_2008, kuhn_feature_2019}). Additionally, we applied a simple \texttt{l2} regularization technique with $\lambda=5\times10^{-4}$ (e.g. \citealt{phaisangittisagul_analysis_2016, van_laarhoven_l2_2017}).

In order to increase the accuracy of the method and provide error estimates, we employ a technique known as deep ensembling (\citealt{lakshminarayanan_simple_2017}). This method leverages the fact that each neural network is independent of the other.  We train ten individual networks with the same architecture but with different weight initializations. We then apply each network to the test data individually. We average the classification probabilities and allow the standard deviation to act as an uncertainty estimate. 

\subsection{SITELLE Data}
SITELLE observations of the Southwest Field of M33 led by P.I. Laurie Rousseau-Nepton were taken during the Queued Service Observation Period 18B (Program 18BP41). The galaxy was imaged in the three primary filters: SN1, SN2, and SN3. Both the SN1 and SN2 observations were taken at a spectral resolution $R\sim1000$, while the SN3 observation was taken at a spectral resolution $R\sim5000$. Although we do not analyse the data from the following two galaxies in detail in this paper, we demonstrate the feasibility of applying the network to two additional SIGNALS galaxies: M95 (Program 19A) and NGC4214 (Program 18A).
These observations represent a subsample of the SIGNALS survey. We note that the authors are members of the SIGNALS science team. The   raw   data   were   reduced   and   calibrated   using   SITELLE’s   Level-1 reduction software,   ORBS   (version  3.1.2 \citealt{martin_orbs_2012}). 
\subsection{Dereddening}\label{sec:red}
Since we are using emission-line ratios spanning a large range of wavelengths, we must include the effects of dust attenuation by reddening the spectra (\citealt{calzetti_dust_1994}; \citealt{buat_star_1996}; \citealt{pettini_rest-frame_2001}). We calculate dereddening following the standard procedure of postulating an effective dust screen attenuation to obtain the intrinsic emission-line fluxes ($F_{0,\lambda}$),
\begin{equation}
    F_{0,\lambda} = F_{\text{obs}, \lambda} \, e^{\tau_\lambda} = F_{\text{obs}, \lambda} \, e^{\tau_V \, q_\lambda},
\end{equation}
where $F_{\text{obs}, \lambda}$ is the observed flux, $\tau_\lambda$ is the optical depth at a given wavelength, $\tau_V$ is the optical depth in the $V$-band, and the shape of the dust attenuation curve is parametrized by $q_\lambda \equiv \tau_\lambda/\tau_V$.
We adopt the \citet{cardelli_relationship_1989} attenuation law with a total-to-selective extinction $R_V = 3.1$. 
We use the Balmer decrement, defined as $B_d = F_\mathrm{obs,H\alpha}/F_\mathrm{obs,H\beta}$, to calculate $\tau_V$:
\begin{equation}
    \tau_V = \frac{1}{q_{H\beta}-q_{H\alpha}}\ln{\frac{B_d}{B_{d,in}}},
\end{equation}
where $B_{d,in}$ is the intrinsic Balmer decrement. We assume this value to be 2.87, which is appropriate for the Case B, an electronic temperature of 10 000 K and low density (e.g. \citealt{osterbrock_astrophysics_1989}). Dereddening is applied to the line ratios post calculation by the neural network.

\section{Results} \label{sec:res}
\subsection{Optimal Synthetic Data}
In this section we discuss the primary results of the paper and address the efficacy of the ANN when applied to optimal SITELLE HII regions sythethic spectra in order to predict the following strong emission-line ratios : \nii{}$\lambda$6548/\nii{}$\lambda$6583, [{\sc S\,ii}]$\lambda6731$/[{\sc S\,ii}]$\lambda6716$, ([{\sc S\,ii}]$\lambda6716$+[{\sc S\,ii}]$\lambda6731$)/H$\alpha$,
[{\sc N\,ii}]$\lambda$6583/H$\alpha$, H$\alpha$/H$\beta$,
 [{\sc N\,ii}]$\lambda$6583/([{\sc S\,ii}]$\lambda6716$+[{\sc S\,ii}]$\lambda6731$), [{\sc O\,iii}]$\lambda$5007/H$\beta$, ([{\sc O\,ii}]$\lambda$3726+\oii{}$\lambda$3729)/[{\sc O\,iii}]$\lambda$5007, and\\
 ([{\sc O\,ii}]$\lambda$3726+\oii{}$\lambda$3729)/H$\beta$. 

\begin{figure*}
    \centering
    \includegraphics[width=1.0\textwidth]{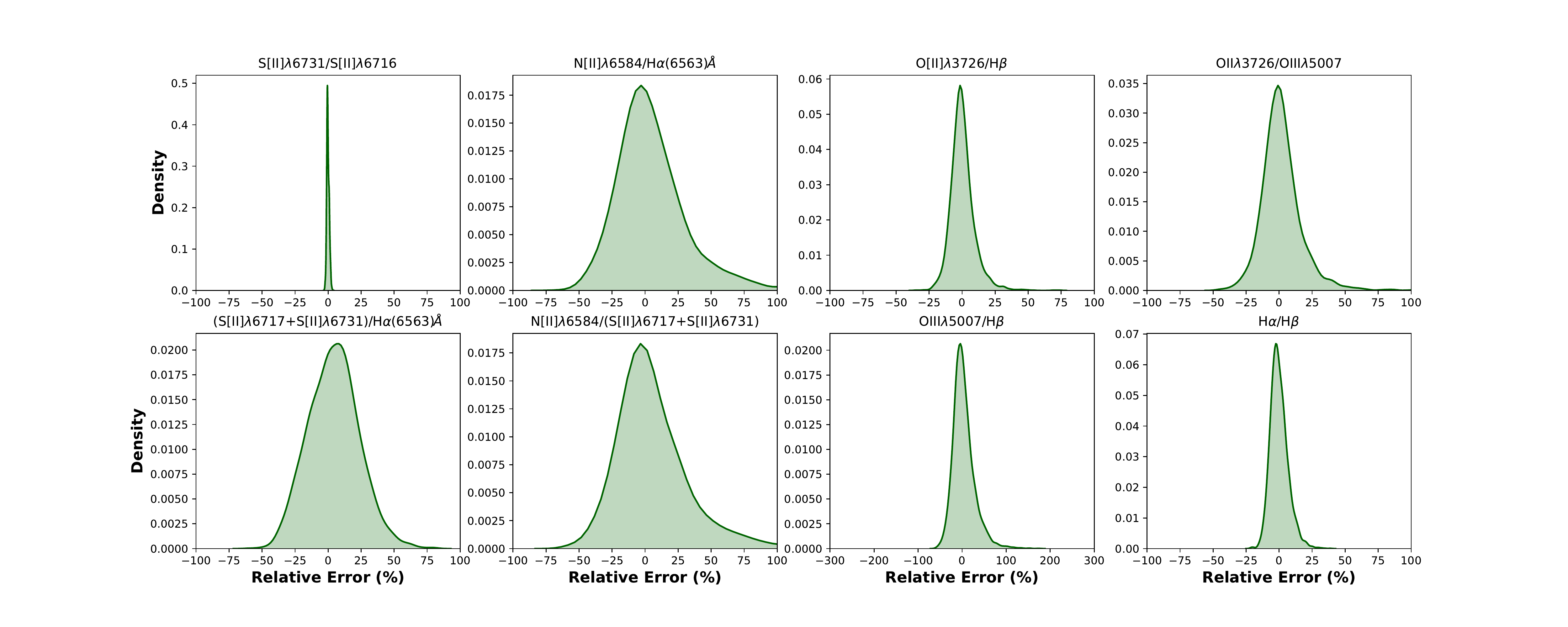}
    \caption{Density plots of line ratio relative errors calculated using the ANN described in $\S$\ref{sec:ann}. We note that the errors follow an approximately Gaussian 
    distribution with a non-negligible positive skew.}
    \label{fig:test-set-res}
\end{figure*}

We train and validate the network using the synthetic data described in $\S$\ref{sec:syn}. The mean absolute errors ($\sum_i^N \texttt{abs}(y_{true_i}-y_{pred_i})$; N is the number of training and validation input) on the training and validation are 0.0555 and 0.1207, respectively. The algorithm is applied to the test set; Figure \ref{fig:test-set-res} shows the relative errors achieved by the network when recovering each emission-line ratio. We note that all line-ratios recovered from SN3 have a low residual standard deviation; this is attributed to the higher spectral resolution in SN3 compared to SN1 and SN2.
All error plots reveal approximately Gaussian error distributions with a positive skew.  
In order to validate these results using the standard fitting techniques, we use the \texttt{ORB} routine \texttt{fit.fit\_lines\_in\_spectrum}. We individually fit each filter; however, within a given filter, we fit all emission-lines simultaneously and supply the routine with the correct velocity and broadening parameters initially in order to retrieve the best possible fits. The lines were fit with a \texttt{sincgauss} function (\citealt{martin_optimal_2016}). Table \ref{tab:ORBvsANN} displays the comparison between the relative errors obtained using standard fitting routines and our ANN on the training set. In these conditions, the network outperforms the standard method for line ratios in all three filters. It is important to note that the relatively high errors in the fits are largely due to signal-to-noise effects (see $\S$\ref{sec:snr} for the discussion). Appendix \ref{app:snr_resid} contains the residuals binned by signal-to-noise for both the neural network and standard fits. These plots illustrate that the neural network is capable of outperforming a fitting routine in the low SNR (signal-to-noise) regime when compared within the training set environment.  
Importantly, the H$\alpha$/H$\beta$ errors have the potential to be reduced significantly which will lead to higher fidelity dereddening estimates.

\begin{table*}[]
    \centering
    \begin{tabular}{|c|c|c|c|c|c|c|c|c|}
         \hline
         Fitting Procedure &  [{\sc S\,ii}]$\lambda6731$/[{\sc S\,ii}]$\lambda6716$ & [{\sc S\,ii}]/H$\alpha$  & [{\sc N\,ii}]/H$\alpha$ & [{\sc N\,ii}]/[{\sc S\,ii}] & [{\sc O\,iii}]/H$\beta$ & [{\sc O\,ii}]/H$\beta$ & [{\sc O\,ii}]/[{\sc O\,iii}] & H$\alpha$/H$\beta$ \\
         \hline \hline 
         \texttt{ORB } & 426.68\% & 17.87\% & 36.67\% & 163.56\% & 190.89\% & 96.30\% & 558.75\% & 90.90\% \\
         \hline
         \texttt{ORB (SNR>10)} & 42.51\% & 18.74\% & 19.06\% & 26.67\% & 35.82\% & 55.43\% & 100.07\% & 20.14\% \\
         \hline
         \texttt{ANN} & 0.94\% & 19.18\% & 30.242\% & 31.67\% & 24.68\% & 8.99\% & 15.36\% & 6.75\% \\ 
         \hline
    \end{tabular}
    \caption{Standard deviation calculations for relative errors of strong emission-line ratios. The top row reports values calculated using the standard \texttt{ORB} routine while the second row calculates the same values for test set spectra with a signal-to-noise ratio greater than ten; comparatively, the bottom row contains values obtained using our ANN. \sii{} refers to the addition of the \sii{} doublets: \sii{}$\lambda$6713 and \sii{}$\lambda$6731; similarly, \oii{} refers to the addition of the two primary \oii{} lines: \oii{}$\lambda$3726 and \oii{}$\lambda$3729. Note that \oiii{} refers only to a single line: \oiii{}$\lambda$5007.
    As expected, the values calculated can be significantly elevated because the test set contains spectra in the extremely low signal-to-noise regime. We discuss the effects of signal-to-noise in detail in $\S$\ref{sec:snr}.}
    \label{tab:ORBvsANN}
\end{table*}

\subsection{Noise Classification}
In this section we evaluate the network used to classify a spectrum as noisy or clean; we define a spectrum clean if the SNR of all strong emission-lines is above a certain pre-determined threshold -- in this case 5\% that of H$\alpha$. This value was determined by running our network on the SN3 spectra of varying thresholds (from 1\%--20\%). The results indicated that below 5\%, the ability of the network to recover the line ratios becomes inhibited.

Synthetic spectra for all filters were created in a method identical to that described in $\S$ \ref{sec:syn} except that we removed the requirement that all strong emission-lines have an amplitude equal to at least 12\% that of H$\alpha$. Spectra were classified as noisy if any emission-line had an amplitude less than 5\% that of H$\alpha$; otherwise, the spectrum was classified as noiseless. In spectra where a single strong emission-line amplitude was below the chosen threshold, several other lines were also beneath the threshold; thus, this constraint accurately categorizes data into noisy and noiseless. We created 1,000 noisy and 1,000 noiseless spectra with signal-to-noise ratios of H$\alpha$ varying between 5 and 30. 


We use a Decision Tree Classifier, but in order to reduce bias and probabilistic errors, we aggregate several trees into a Random Forest. The data were randomly shuffled; 90\% were set aside for training, and 10\% for testing. Using 10 estimators (or 10 decision trees), the Random Forest Classifier reports 100\% accuracy in classifying the two spectral types resulting in a diagonal confusion matrix. Although reaching perfect accuracy is relatively rare in machine learning scenarios, the well-defined and simple task allow the Random Forest Classifier to achieve 100\% accuracy. Furthermore, if instead 70\% of the data is taken for training, the power of the algorithm allows us to simply increase the number of estimators in order to obtain the same perfect accuracy. A future paper will dig deeper into the Noise classification problem.


\section{Discussion}

\subsection{Verification of the Network}
Although neural networks trained on synthetic data are notoriously difficult to verify, we explore several methods to test whether the network is accurately learning the line ratios and is portable to real data (e.g. \citealt{bishop_novelty_1994}; \citealt{krogh_neural_1995}). We apply the following three techniques: k-fold cross-validation, tracking the static \nii{}$\lambda$6548/\nii{}$\lambda$6583 ratio, and a saliency map. 

We apply a standard k-fold cross-validation by partitioning our synthetic data set into ten equal-sized allotments. We then train and validate the network with nine of the ten partitions allowing the final partition to serve as the test set. This process is repeated until each partition has been served as a test set (e.g. \citealt{bengio_no_2004}; \citealt{picard_cross-validation_1984}). This method validates the network's estimates by testing the network for overfitting. Overfitting occurs when the network learns the training set but is unable to generalize to new data. If the network is susceptible to overfitting, then the final accuracy scores of the k-fold cross-validation model will differ significantly. A k-fold cross-validation analysis of our network indicates no overt overfitting in the model since the accuracy are all within several percentage points of one another (\citealt{molinaro_prediction_2005}; \citealt{cawley_over-tting_2010}).

\begin{figure}
    \centering
    \includegraphics[width=0.48\textwidth]{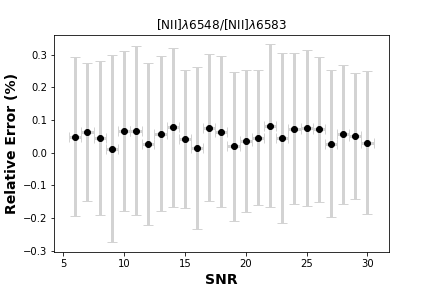}
    \caption{Relative error as a function of signal-to-noise for the \nii{} doublet: \nii{}$\lambda$6548/\nii{}$\lambda$6583. The black dots represent the median relative error in a given signal-to-noise bin while the grey bars are the 1-$\sigma$ errors associated with a given bin.}
    \label{fig:NII-doublet}
\end{figure}

Additionally, we estimate the line ratio \nii{}$\lambda$6548/\nii{}$\lambda$6583 which is expected to be constant throughout all emission-line nebula and is frequently set to 3 in fitting code (e.g. \citealt{schirmer_sample_2013}). This is reflected in the synthetic data set. Figure \ref{fig:NII-doublet} reveals that the calculated \nii{} doublet ratio relative error between the true and network-estimated value is between 0 and -1\% while the standard deviation is approximately 0.27\%. Therefore, the network accurately replicates the static relation between the \nii{} lines. 

\begin{figure*}
    \centering
    \includegraphics[width=0.98\textwidth]{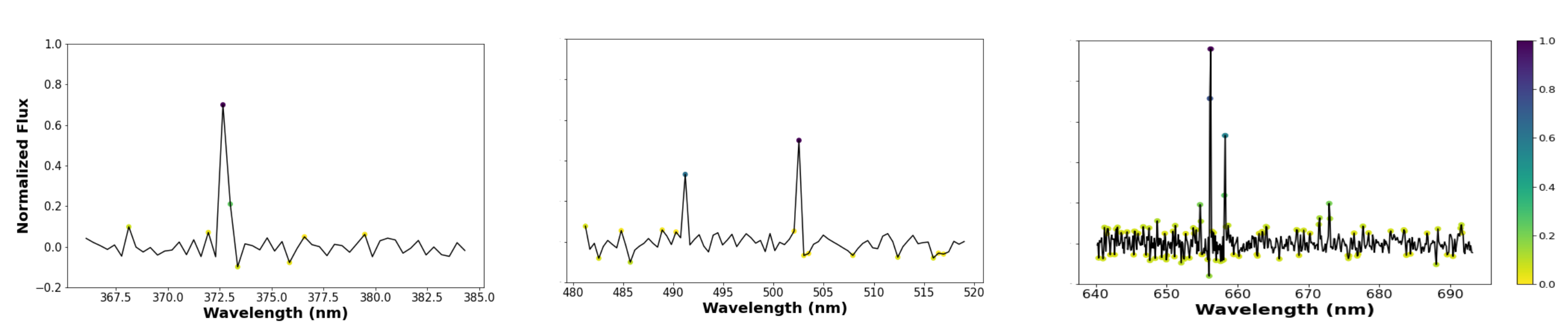}
    \caption{Activation Map of the SN1, SN2, and SN3 filters using the reference spectrum. The relative weights are centered on emission line peaks and surrounding regions reflecting the importance of amplitude and side-lobes on flux ratio estimations. The signal-to-noise of this sample spectrum is approximately 20.}
    \label{fig:activation}
\end{figure*}

Subsequently, we calculate the saliency map of the network. In order to calculate the saliency map, we calculate the derivative of the input with respect to the output of the neural network: $\frac{\partial [input]}{\partial[output]}$.
By multiplying the gradients together, we are left with our desired partial derivative (e.g. \citealt{simonyan_deep_2014}). We normalize the values to unity in order to compare their relative importance with ease. In this manner, input nodes with a saliency value of zero do not affect the neural network, while input nodes with a saliency value of 1 have the most importance in affecting the neural network's estimation. In Figure \ref{fig:activation}, we show the saliency values plotted over a reference spectrum with a signal-to-noise of 20; we note that saliency values less than 0.05 are not included in the figure. The saliency map clearly shows that the network prioritizes the amplitude of the emission lines. However, the map also reveals the importance of the regions adjacent to the emission-lines and the areas of the spectrum in between emission-lines. This reflects the importance of the sidelobes of the ILS on the flux ratio estimates.

\subsection{Effects of Varying Signal-to-Noise}\label{sec:snr}
Signal-to-noise constrains the efficacy of traditional fitting methods (e.g. \citealt{campbell_stellar_1986}; \citealt{endl_kea_2016}). In this section, we compare the effects of signal-to-noise on the standard fitting techniques implemented in \texttt{ORB} compared to the artificial neural network described in this paper. 

In order to study the effects of the signal-to-noise on the efficacy of line ratio estimates, we bin the line ratio residuals as a function of signal-to-noise. A signal-to-noise bin is created at each integer value of the sampled signal-to-noise used to create the synthetic data (5-030). Residuals were calculated by taking the estimated value and subtracted the ground truth and dividing that value by the ground truth value; the value was then multiplied by 100 to make it a percentage. We then removed all outliers; outliers are defined as residual values more than 3-$\sigma$ off the median value. We then calculated the median value of the remaining set of residuals in each signal-to-noise bin. Errors were calculated as the 1-$\sigma$ deviations from the median.
The plots are shown in appendix \ref{app:snr_resid}. Figure \ref{fig:ml_snr} demonstrates that the residuals do not change as a function of the signal-to-noise when calculated by the neural network. Conversely, the residuals and their associated errors are greatly reduced in high signal-to-noise regimes ($R>20$) when calculated using standard fitting techniques.

\subsection{Application to M33}
Having demonstrated the feasibility of using a neural network to estimate strong emission-line ratios, we apply our methodology to the Southwest field of M33 studied in our previous article (\citealt{rhea_machine_2020}). This field contains several previously identified emission region types (classic \hii{} regions, supernova remnants, and planetary nebulae; e.g. \citealt{zaritsky_kinematics_1989}; \citealt{hodge_hii_1999}; \citealt{viallefond_hii_1986}); additionally, this field is a SIGNALS target. All fits (both from the algorithm developed here and \texttt{ORCS}) were run using a computing server located at the CFHT headquarters in Waimea, Hawaii named \textit{iolani}. The server has 2 Intel \texttt{XEON E5-2630 v3} CPUs operating at 2.40GHz with 8 cores each. The configuration also has 64 GB of RAM available for computing purposes. 

To compare our results with those from the \texttt{ORCS} fitting pipeline, we fit each cube seperately using \texttt{ORCS}. We use the \texttt{fit\_lines\_in\_region} function to fit the strong emission-lines present in a given filter. Within a filter, the lines are fit simultaneously with a single sinc function convolved with a Gaussian which returns each line's flux, velocity, and broadening (\citealt{martin_calibrations_2017}). \texttt{ORCS} fits the observed data to a sinc function convolved with a Gaussian using the Levenberg-Marquardt least squares optimization algorithm. Velocity and broadening priors were determined by fitting a binned (8$\times$8) data cube. The final unbinned fits ($\sim$ 4 million spaxels) of SN3 took approximately 11 days, SN2 took 5.5 days, and SN1 took 1.8 day. We note that these calculations are done in parallel. Additionally, we applied the trained neural network to the cube. By comparison, the machine learning methodology requires approximately 18 computing hours, non-parallelized, to complete the entire cube (all three filters). This represents a reduction in computation time by approximately 2 orders of magnitude. Additionally, we applied the network outlined in \cite{rhea_machine_2020} to calculate the velocity and broadening in order to have a more robust timing analysis. We report that the addition of this network did not add any significant amount of time to the overall fitting of the data cube. These results reinforce the notion introduced in \cite{rhea_machine_2020} that by taking a machine learning approach to the analysis of IFU data cubes we can greatly reduce the time required for computation while retaining the accuracy of the results with a proper training. 

Although there does not exist a known ground-truth value for these line ratios, we calculate the residuals between the machine learning predictions and standard \texttt{ORCS} fits in order to quantify the accuracy of the algorithm. Normalized residual plots for each line ratio can be found in appendix \ref{app:line_res}. A visual inspection of the residual plots reveal that the machine learning algorithm returns similar values to those calculated by \texttt{ORCS}. Results deviate most strongly in regions which we show in \cite{rhea_machine_2020} to be best described by multiple emission profiles, regions identified as non-\hii{} regions, and those with a low signal-to-noise ratio. 
We note, however, that the normalized residual plots (Figures \ref{fig:res_plots1} and \ref{fig:res_plots2}) show general agreement between the standard \texttt{ORCS} fits and those calculated by the neural network. Furthermore, discrepancies between the \texttt{ORCS} and neural network estimates illustrate the limitations of such an approach when used as a replacement to global fitting algorithms such as \texttt{ORCS}.
These results further indicate the importance of taking multiple emission profiles into account when modeling -- this is the topic of the following paper in the series. 

\begin{figure}[t!]
    \centering
    \includegraphics[width=0.5\textwidth]{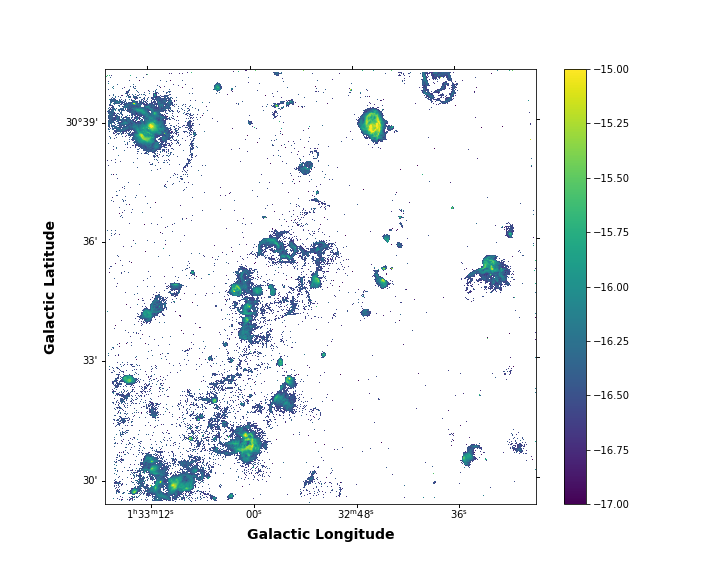}
    \caption{Coadded H$\alpha$ and \nii{}$\lambda6583$ emission map. The image illustrates the density of emission-line regions in the outskirts of M33. All pixels with an H$\alpha$ flux less than $2\times10^{-17}$erg s$^{-1}$ cm$^{-2}$ are masked out which corresponds to a signal-to-noise ratio of approximately 5.}
    \label{fig:DeepM33}
\end{figure}

\section{Conclusions}
Applications of machine learning in astronomy are broad: from the estimation of stellar spectral parameters (e.g., \citealt{fabbro_application_2018}) to the discovery of objects of interest in extensive astronomical surveys (e.g., \citealt{skoda_active_2020}). In this work, we apply an artificial neural network to combined-filter (SN1, SN2, and SN3) SITELLE data representing typical SIGNALS large program observations. The network is designed to calculate important emission-line ratios for \hii{}-like regions which are present in the primary SITELLE filters. We train, validate, and test the algorithm using synthetic data created with the \texttt{ORBS} software package. We adopt physically-derived line amplitudes from the Million Mexican Model Database (\citealt{morisset_virtual_2015}). Our results indicate that the network can potentially constrain the line ratios with greater precision than the standard line fitting technique implemented in \texttt{ORCS} if the source spectral properties are well represented in the training set. To demonstrate the applicability of the method beyond synthetic data, we apply the network to the Southwest field of M33. Timing analysis indicates that the network can analyze the entire cube approximately \carter{100} times faster than the standard methods.

These results not only have an impact on the computational aspects of line ratio calculations, but they also carry scientific implications. Although our knowledge of galactic dynamics has expanded considerably over the past several decades, spectroscopic conclusions are restricted by the rigor of the fitting schemes employed and precision of the results. In this paper, we have demonstrated that machine learning algorithms can considerably increase the precision of emission line ratios in both low and high signal-to-noise regimes. This has profound implications on the study of these regions since it will allow stricter categorization using methods such as line-ratio diagnostics in conjunction with BPT diagrams (e.g.  \citealt{baldwin_classification_1981}; \citealt{kewley_host_2006}; \citealp{kewley_understanding_2019}). These methods require concise measurements in order to accurately categorize the emission-region type and break any model degeneracies. 

Following up on the success of our first report, the work presented here represents the second article in a series of articles covering the application of machine learning algorithms to SITELLE data cubes. Our results have been encouraging in mapping out emission line ratios in pixels dominated by \hii{} region emission serves as a proof-of-concept that using machine learning to identify line fluxes is a viable methodology. We note this work is not meant to be a replacement to global line fitting algorithms. Identifying regions containing multiple, blended emission components, as well as multiple sources of emission with spectral features not represented in a training set remains to be explored. 
Additionally, example code can be found at \url{https://github.com/sitelle-signals/Pamplemousse}.

\acknowledgments
The authors would like to thank the Canada-France-Hawaii Telescope (CFHT) which is operated by the National Research Council (NRC) of Canada, the Institut National des Sciences de l'Univers of the Centre National de la Recherche Scientifique (CNRS) of France, and the University of Hawaii. The observations at the CFHT were performed with care and respect from the summit of Maunakea which is a significant cultural and historic site.
C. R. acknowledges financial support from the physics department of the Universit\'e de Montr\'eal, the MITACS summer scholarship program, and the IVADO doctoral excellence scholarship.
J. H.-L. acknowledges support from NSERC via the Discovery grant program, as well as the Canada Research Chair program.
NVA acknowledges support of the Royal Society and the Newton Fund via the award of a Royal Society--Newton Advanced Fellowship (grant NAF\textbackslash{}R1\textbackslash{}180403), and of Funda\c{c}\~ao de Amparo \'a Pesquisa e Inova\c{c}\~ao de Santa Catarina (FAPESC) and Conselho Nacional de Desenvolvimento Cient\'{i}fico e Tecnol\'{o}gico (CNPq).

\appendix

\section{Signal-to-Noise and Residuals}\label{app:snr_resid}
\begin{figure}
    \centering
    \includegraphics[width=1.0\textwidth]{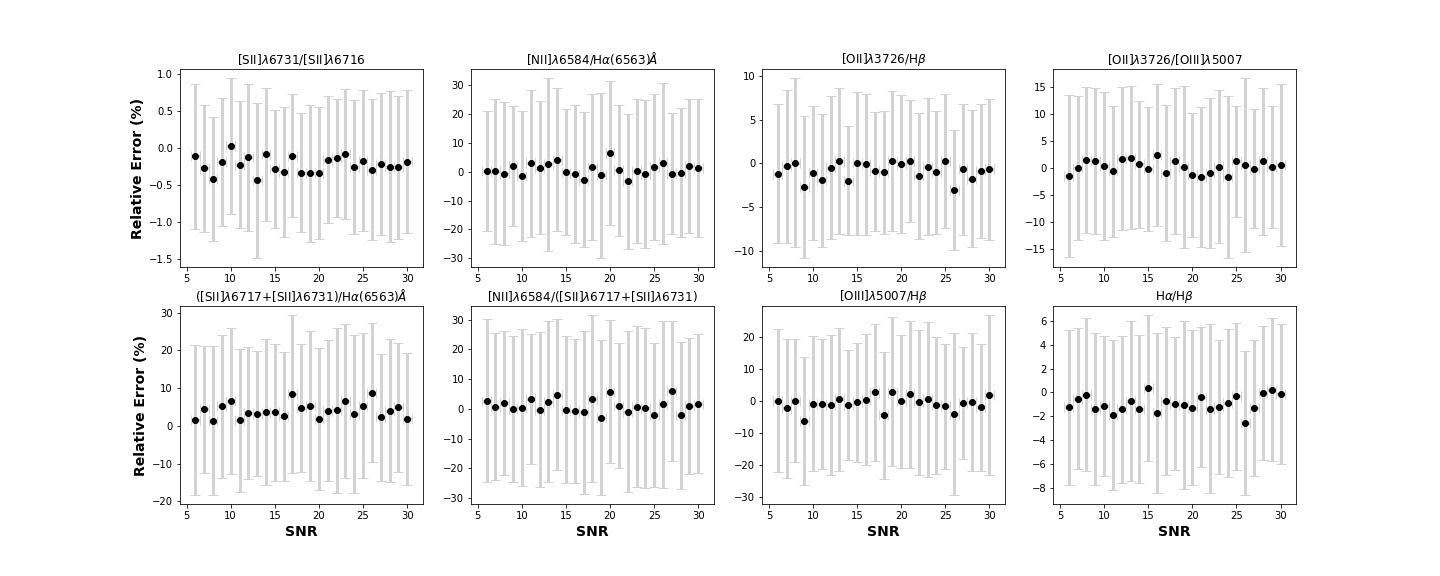}
    \caption{Signal-to-noise ratio vs relative errors for the estimations obtained on the test set using the artificial neural network described in this paper. The signal-to-noise bins were taken at integer intervals. The black dots are the mean residuals. The grey bars represent the 1-sigma errors in a given signal-to-noise bin.}
    \label{fig:ml_snr}
\end{figure}

\begin{figure}
    \centering
    \includegraphics[width=1.0\textwidth]{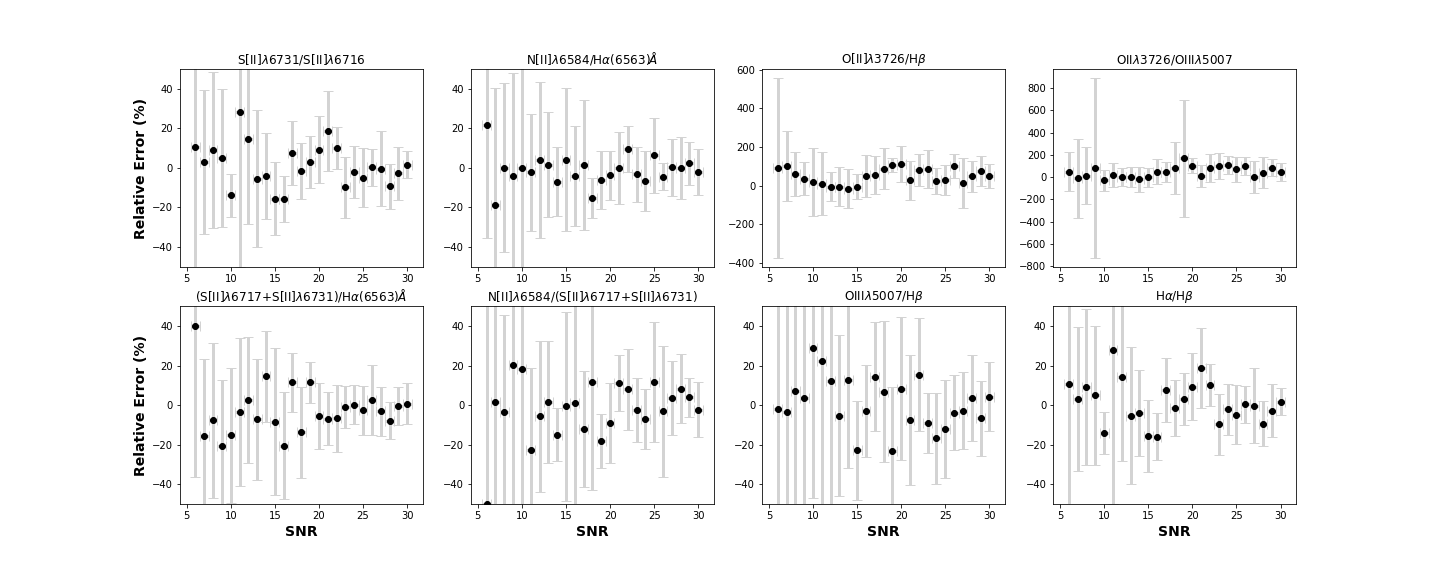}
    \caption{Signal-to-noise ratio vs relative errors for the estimations obtained on the test set using the software package \texttt{ORBS}. The black dots are the mean residuals. The signal-to-noise bins were taken at integer intervals. The grey bars represent the 1-sigma errors in a given signal-to-noise bin.}
    \label{fig:orcs_snr}
\end{figure}

In this section we display the signal-to-noise vs residual plots described in $\S$\ref{sec:snr}. Figure \ref{fig:ml_snr} displays the signal-to-noise vs residual plots for each line ratio calculated using the machine learning algorithm described in $\S$\ref{sec:ann}. Figure \ref{fig:orcs_snr} shows the signal-to-noise vs residual plots for each line ratio calculated using \texttt{ORB} as described in $\S$\ref{sec:res}.

\section{Line Ratio Residual Plots}\label{app:line_res}
\begin{figure*}
\gridline{\fig{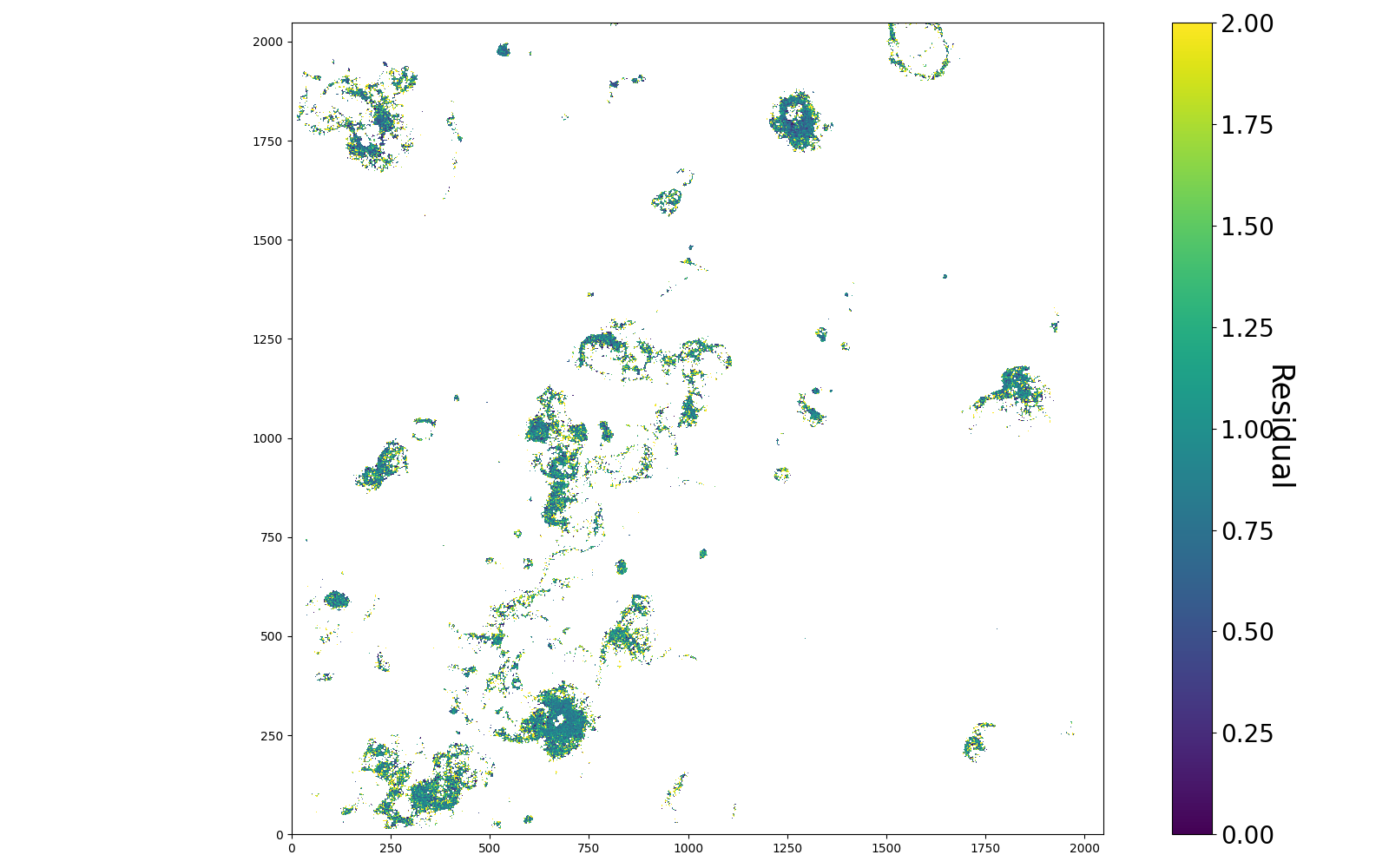}{0.5\textwidth}{(a)\sii{}$\lambda$6731/\sii{}$\lambda$6716}
          \fig{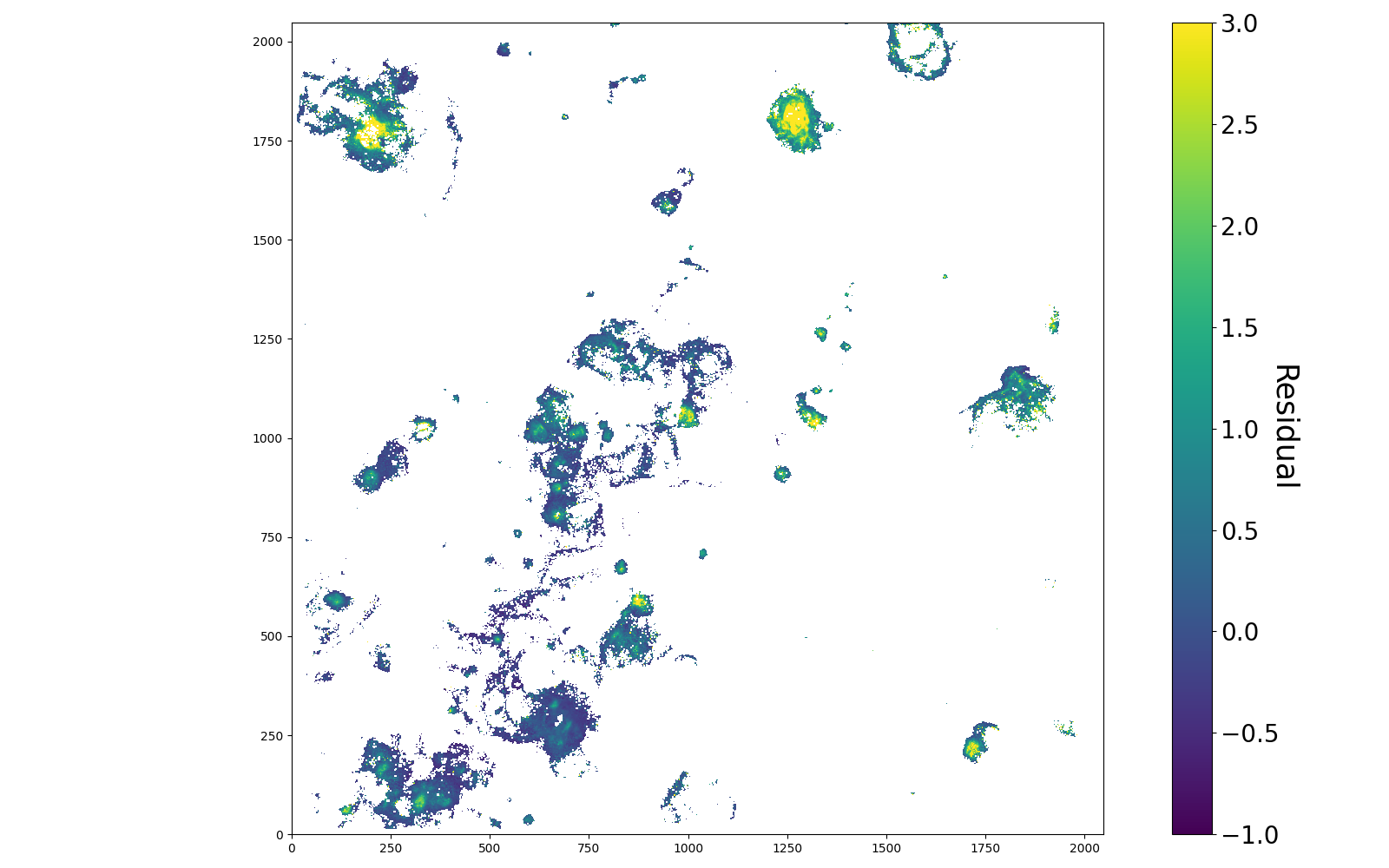}{0.5\textwidth}{(b)(\sii{}$\lambda$6731+\sii{}$\lambda$6716)/H$\alpha$}
          }
\gridline{\fig{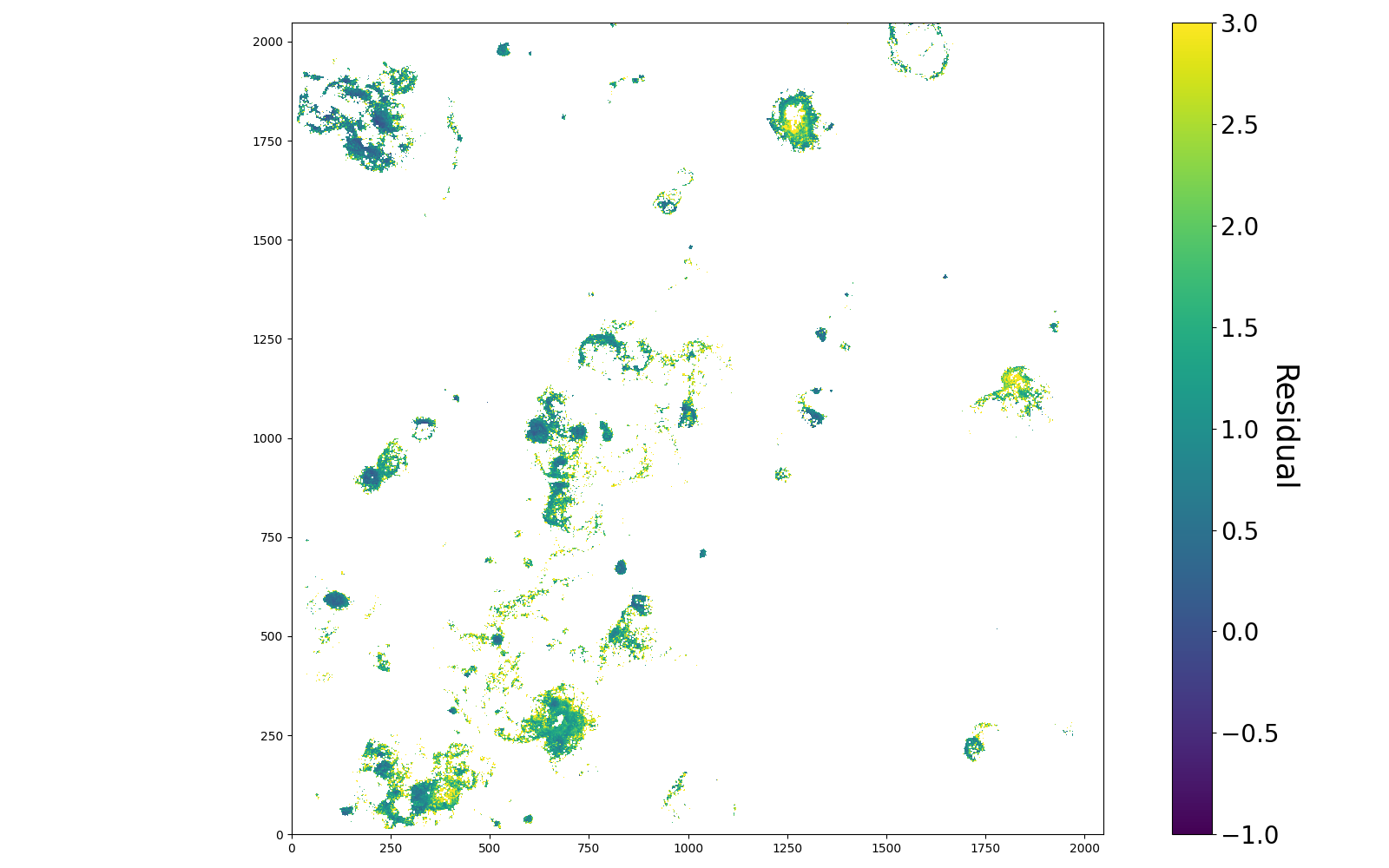}{0.5\textwidth}{(c)\nii{}$\lambda$6583/(\sii{}$\lambda$6731+\sii{}$\lambda$6716)}
          \fig{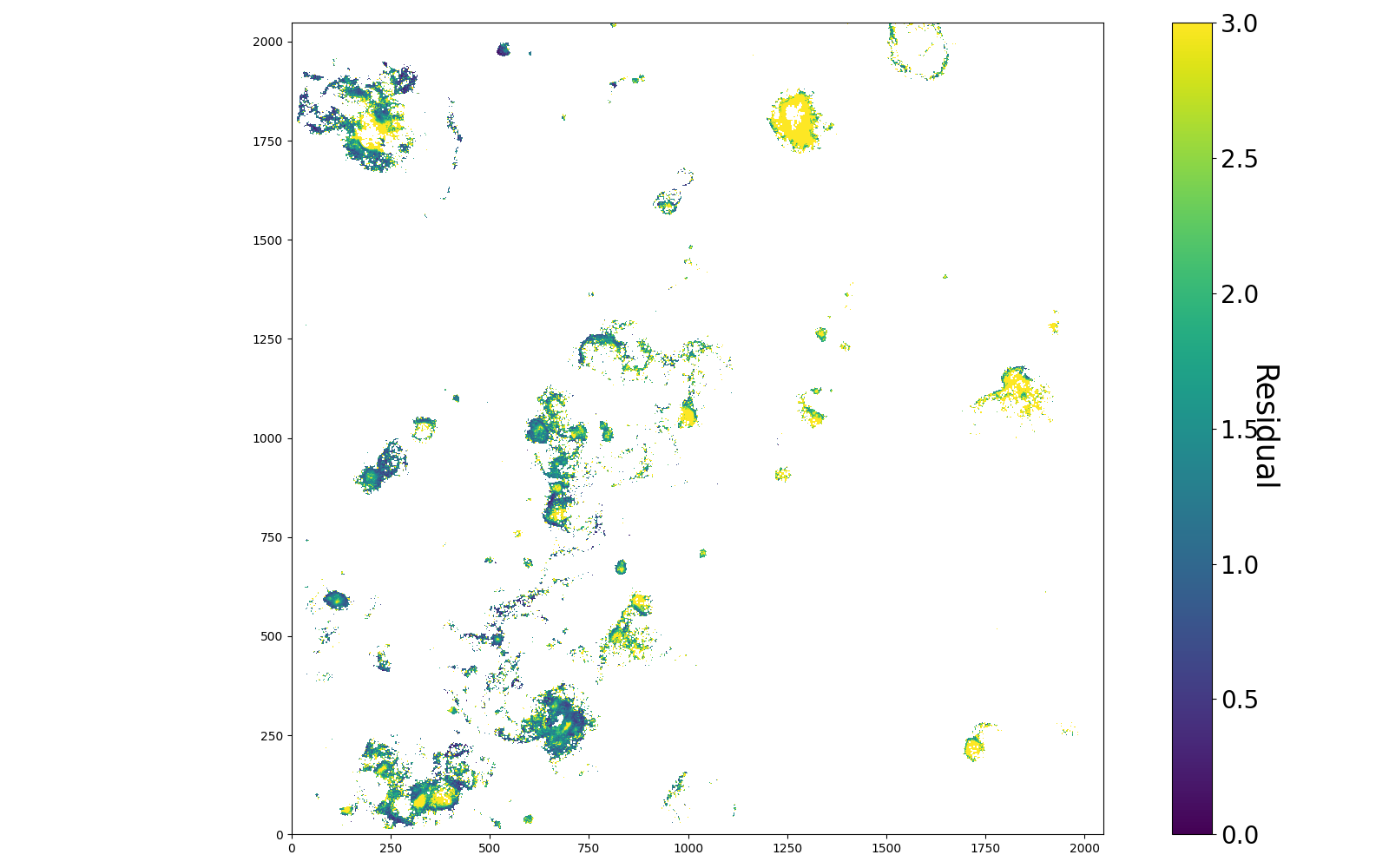}{0.5\textwidth}{(d)\nii{}$\lambda$6583/H$\alpha$}
          }          
\gridline{\fig{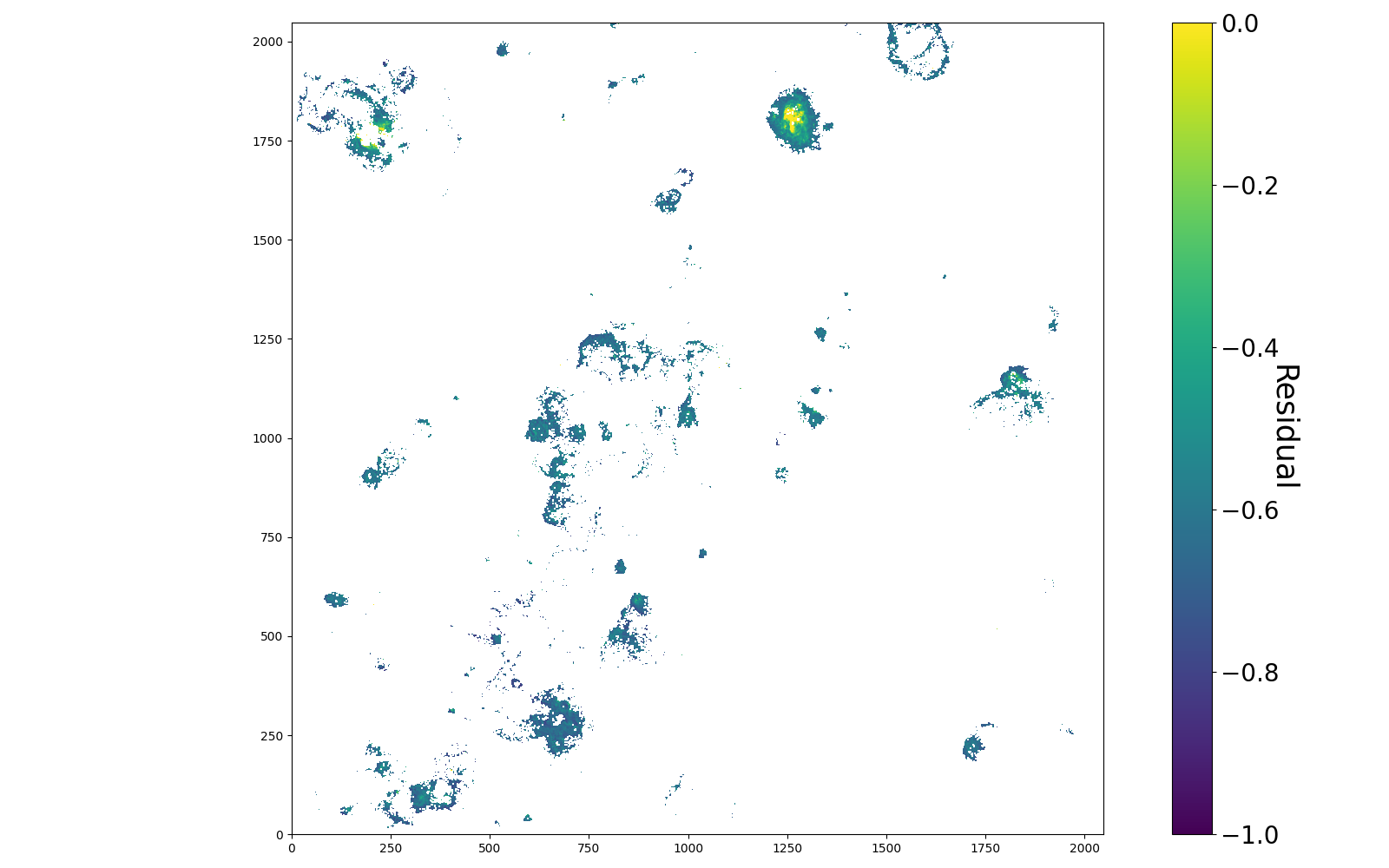}{0.5\textwidth}{(e)(\oii{}$\lambda$3726+\oii{}$\lambda$3729)/H$\beta$}
          \fig{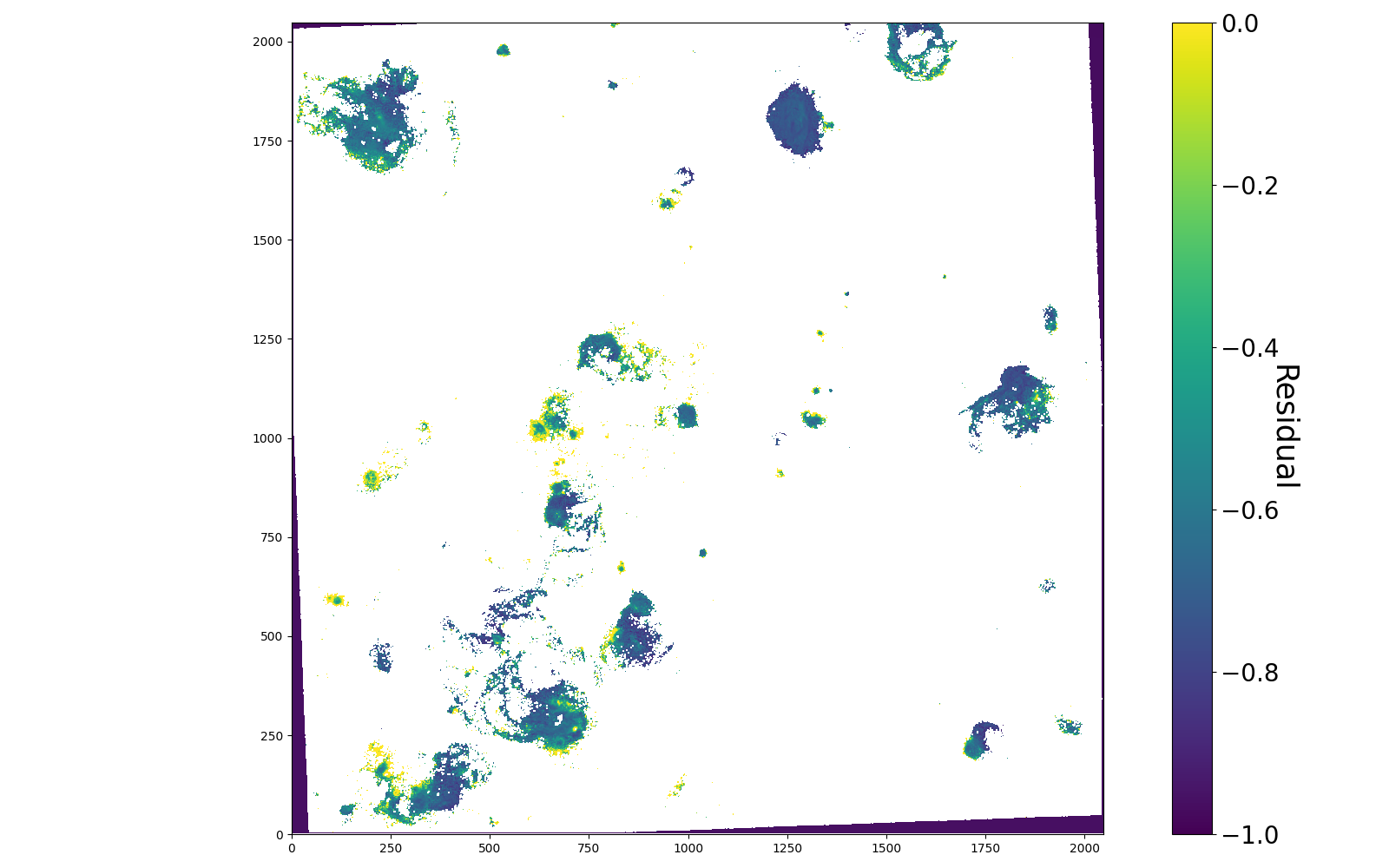}{0.5\textwidth}{(f)\oiii{}$\lambda$5007/H$\beta$}
          }
\caption{Residual Plots created by taking the \carter{difference} between the \texttt{ORCS} fits and the values calculated by the artificial neural network for the Southwest Field of M33 normalized by the \texttt{ORCS} fit values. As discussed in the text, regions with large discrepancies between the ORCS and ANN fits are generally either not classic \hii{} regions or are described best by multiple components.}
\label{fig:res_plots1}
\end{figure*}

\begin{figure*}
\gridline{\fig{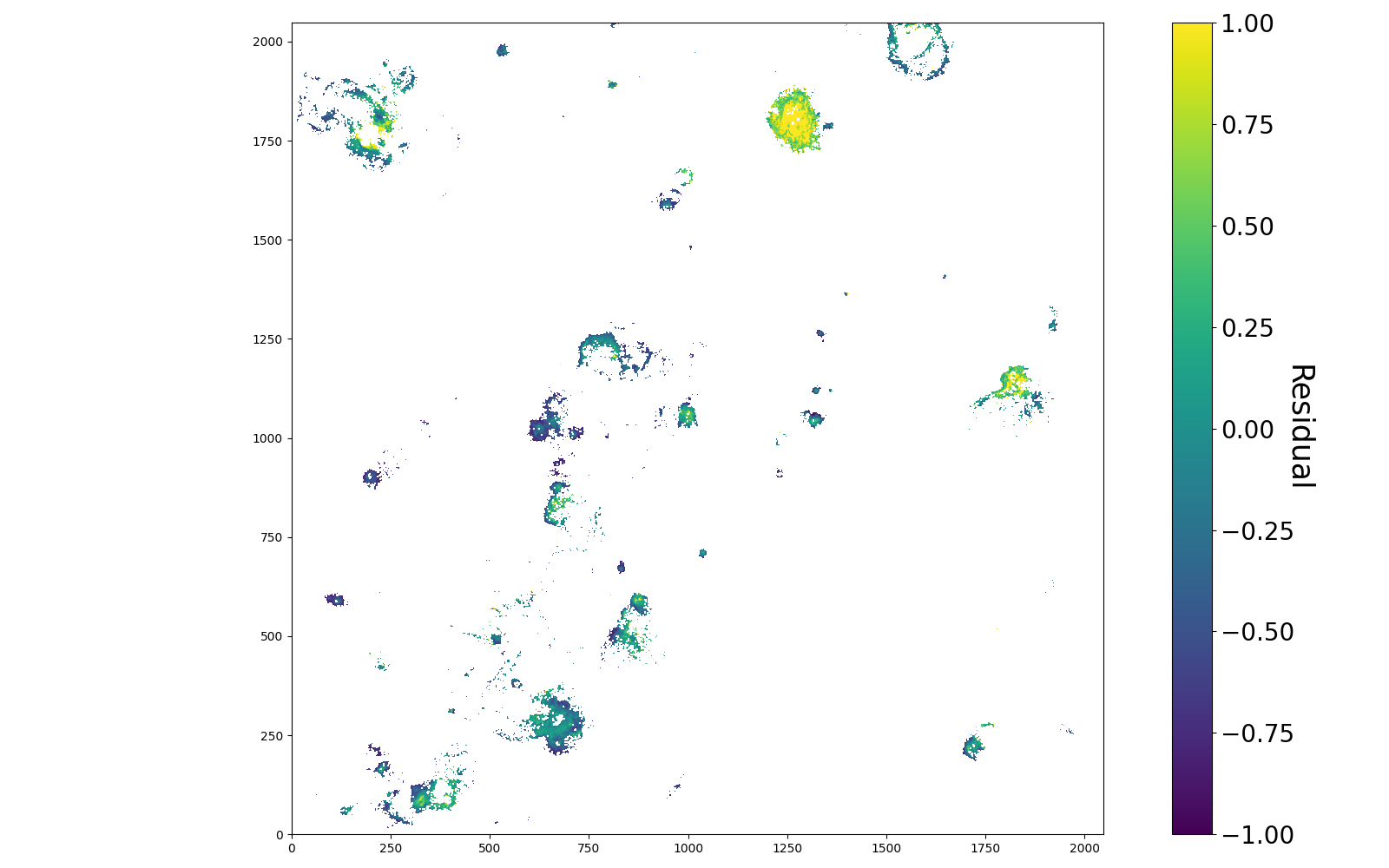}{0.5\textwidth}{(g)(\oii{}$\lambda$3726+\oii{}$\lambda$3729)/\oiii{}$\lambda$5007}
          \fig{residual_s2_s1}{0.5\textwidth}{(h)\sii{}$\lambda$6731/\sii{}$\lambda$6716}
          }
\gridline{
          \fig{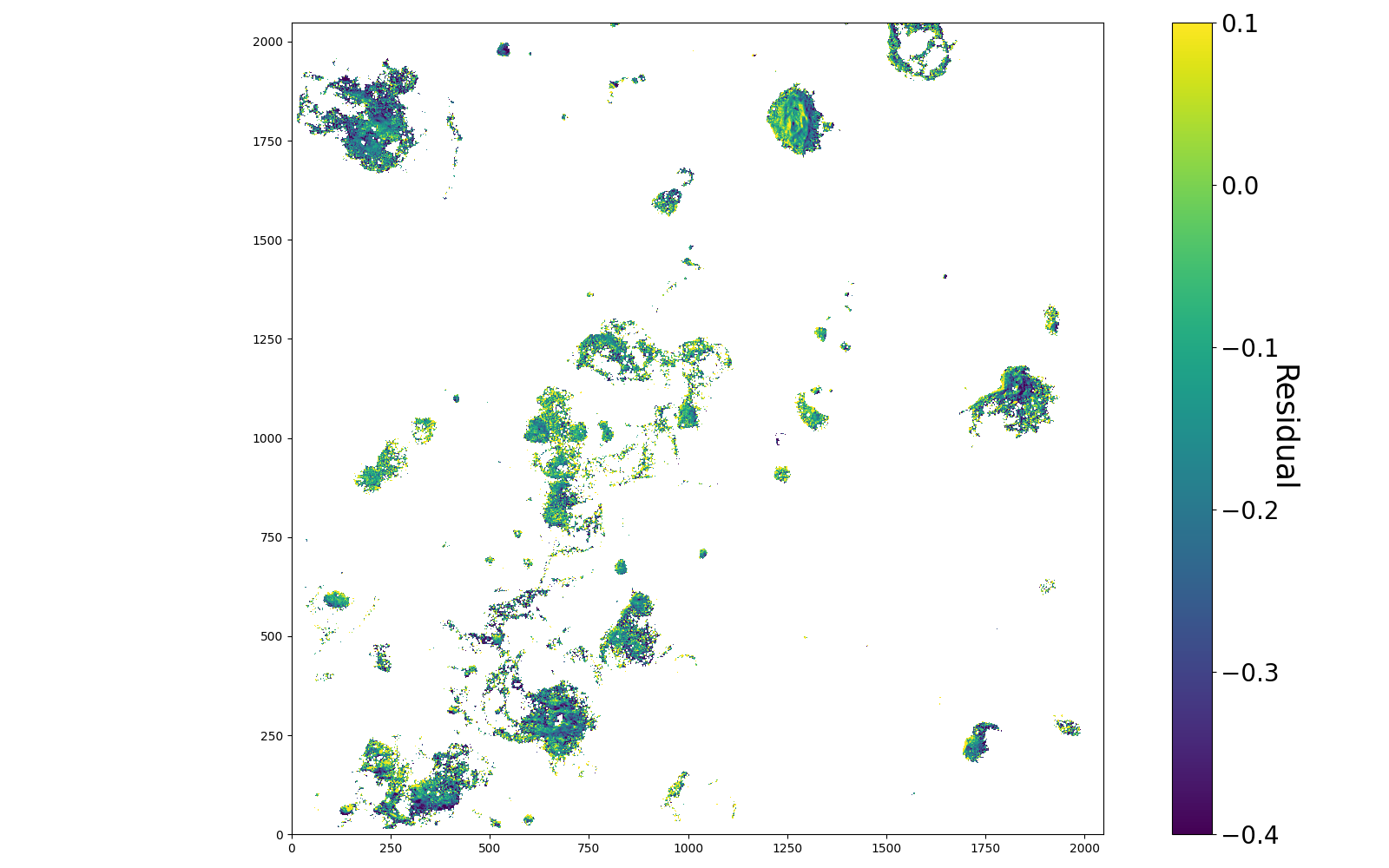}{0.5\textwidth}{(i)H$\alpha$/H$\beta$}
          }
\caption{Extension of Figure \ref{fig:res_plots1}.}
\label{fig:res_plots2}
\end{figure*}

This section contains the line ratio residual plots (ORCS fits - ANN estimates; Figures \ref{fig:res_plots1} and \ref{fig:res_plots2}). The two methods are in agreement in regions found to be best described by a single emission profile for each strong-line (see \cite{rhea_machine_2020} for details); in regions best described by two emission profiles, the results differ significantly. In the subsequent paper, we will explore machine learning techniques to determine whether or not emission regions are best described by a single or double emission profile.

\bibliography{SitelleML2}{}
\bibliographystyle{aasjournal}

-

\end{document}